# Electronic phase diagram of the layered cobalt oxide system, $Li_xCoO_2$ ($0.0 \leq x \leq 1.0$)

T. Motohashi[1,2], T. Ono[2,3], Y. Sugimoto[1], Y. Masubuchi[1], S. Kikkawa[1], R. Kanno[3], M. Karppinen[2,4], and H. Yamauchi[2,3,4]

[1]*Graduate School of Engineering, Hokkaido University, Sapporo 060-8628, Japan*

[2]*Materials and Structures Laboratory, Tokyo Institute of Technology, Yokohama 226-8503, Japan*

[3]*Interdisciplinary Graduate School of Science and Engineering, Tokyo Institute of Technology, Yokohama 226-8502, Japan*

[4]*Laboratory of Inorganic Chemistry, Department of Chemistry, Helsinki University of Technology, FI-02015 TKK, Finland*

Here we report the magnetic properties of the layered cobalt oxide system, $Li_xCoO_2$, in the whole range of Li composition, $0 \leq x \leq 1$. Based on dc-magnetic susceptibility data, combined with results of $^{59}$Co-NMR/NQR observations, the electronic phase diagram of $Li_xCoO_2$ has been established. As in the related material $Na_xCoO_2$, a magnetic critical point is found to exist between $x = 0.35$ and $0.40$, which separates a Pauli-paramagnetic and a Curie-Weiss metals. In the Pauli-paramagnetic regime ($x \leq 0.35$), the antiferromagnetic spin correlations systematically increase with decreasing $x$. Nevertheless, $CoO_2$, the $x = 0$ end member is a non-correlated metal in the whole temperature range studied. In the Curie-Weiss regime ($x \geq 0.40$), on the other hand, various phase transitions are observed. For $x = 0.40$, a susceptibility hump is seen at 30 K, suggesting the onset of static AF order. A magnetic jump, which is likely to be triggered by charge ordering, is clearly observed at $T_t \approx 175$ K in samples with $x = 0.50$ (= 1/2) and 0.67 (= 2/3), while only a tiny kink appears at $T \approx 210$ K in the sample with an intermediate Li composition, $x = 0.60$. Thus, the phase diagram of the $Li_xCoO_2$ system is complex, and the electronic properties are sensitively influenced by the Li content ($x$).

I. INTRODUCTION

The layered cobalt oxide system, $Na_xCoO_2$, has attracted a great deal of attention for various unconventional electronic properties. The crystal of $Na_xCoO_2$ consists of a single-atomic Na layer sandwiched by two $CoO_2$ layers [1]. Cobalt atoms in the $CoO_2$ layer form a triangular lattice that is likely to involve complicated magnetic interactions. It is known that $Na_xCoO_2$ shows a wide range of Na nonstoichiometry [1], and properties of $Na_xCoO_2$ strongly depend on the Na content ($x$). As $x$ decreases, the average valence of cobalt increases toward +4 such that the concentration of magnetic $Co^{IV}$ ($S = 1/2$) gradually increases in a nonmagnetic $Co^{III}$ ($S = 0$) matrix. In $Na_xCoO_2$, antiferromagnetic (AF) ordering is suppressed by geometrical frustration in the cobalt triangular lattice that may cause intriguing electronic/magnetic behaviors. At about $x = 0.7$, $Na_xCoO_2$ exhibits unusually large thermoelectric power and metallic conductivity simultaneously [2]. An increase in $x$ only by 0.05 (*i.e.* at $x = 0.75$) leads to a drastic enhancement in thermoelectric power [3,4] and induces a spin-density-wave state below $T_m = 22$ K [5]. In the lower Na content regime, the $x \approx 0.35$ member readily absorbs water, and the resultant hydrated derivative $Na_xCoO_2 \cdot yH_3O^+ \cdot y'H_2O$ becomes a superconductor with $T_c = 4.5$ K [6]. It was also found [7] that the electronic phase diagram of the $Na_xCoO_2$ system is divided into two distinct regimes: one for the Curie-Weiss metal with $x > 0.5$ and the other for the Pauli-paramagnetic metal with $x < 0.5$, and a charge-ordered state of poor electrical conduction appears at $x = 0.5$, in between these two regimes.

Despite the extensive research in previous works, there still remain several important issues. First, properties have been unknown in the low $x$ regime, *i.e.* $x < 0.25$, due to the difficulty in sample syntheses. The chemical oxidation of $Na_xCoO_2$ ($x \approx 0.7$) was previously examined by means of oxidizing reagents including $Br_2$ and $NO_2BF_4$, but the attainable $x$ was limited down to 0.15 [8]. For deeper understanding of the electronic structure of $Na_xCoO_2$, information on this compositional regime is highly desirable. Particularly, $CoO_2$, the $x = 0$ end member is important since it can be considered as a parent phase of $Na_xCoO_2$. It is remarkable to see how the electronic structure evolves upon electron doping into the "non-doped" $CoO_2$ phase, in order to construct an appropriate theoretical model for the intriguing properties of $Na_xCoO_2$. Second, it has remained unclear whether the triangular $CoO_2$ lattice always exhibits various unconventional properties as reported for $Na_xCoO_2$. Previous structural studies revealed [9-11] that Na ions in $Na_xCoO_2$ tend to form superstructures with characteristic

Na contents. It is suggested that the formation of superstructures obscures the intrinsic nature in the triangular $CoO_2$ lattice, as a Coulomb potential due to long-range Na-ion order may influence the adjacent $CoO_2$ layer.

To address these issues, we focused on the layered cobalt oxide, $Li_xCoO_2$. This compound is regarded as a related material of $Na_xCoO_2$, as both $Li_xCoO_2$ and $Na_xCoO_2$ contain the triangular $CoO_2$ layers in common. It was reported that Li nonstoichiometry in $Li_xCoO_2$ can be widely controlled through electrochemical technique [12,13]. Note that $Li_xCoO_2$ is one of the representative cathode materials for the Li-ion secondary battery owing to its excellent capability of electrochemical de-intercalation of Li. Recently, we successfully obtained single-phase polycrystalline samples of $CoO_2$ [14] and $Li_xCoO_2$ [15] through electrochemical de-intercalation of Li from pure $LiCoO_2$ ($x$ = 1.0) bulks. Approximately 50 – 100 mg of phase-pure sample enabled us to perform precise physical property measurements. Here, we report the magnetic properties of $Li_xCoO_2$ in the whole range of Li composition, $0 \leq x \leq 1$. Based on dc-magnetic susceptibility data, combined with results of our recent $^{59}$Co-NMR/NQR observations [16], the electronic phase diagram of $Li_xCoO_2$ is established. Our result covers a missing area in the phase diagram of $Na_xCoO_2$ and thus contributes to the comprehensive understanding of physics in the triangular $CoO_2$ lattice. In addition, we compare the phase diagram of $Li_xCoO_2$ with that of $Na_xCoO_2$ to highlight similarities and differences between the two systems. The different features in the two systems are discussed from the crystallographic point of view.

## II. EXPERIMENTAL

Polycrystalline samples of $Li_xCoO_2$ and $CoO_2$ ($x$ = 0.0) were synthesized through electrochemical de-intercalation of Li from pristine $LiCoO_2$, as described elsewhere [14,15]. Approximately 100 mg of single-phase $LiCoO_2$ pellet was electrochemically oxidized with a constant current of 0.1 mA (= 0.13 mA/cm$^2$) in an airtight flat cell filled with a nonaqueous electrolyte. No auxiliary agents were added to the bulk pellet to avoid any magnetic noise sources. For each sample, the Li content (or the amount of Li ions to be extracted, *i.e.* 1-$x$) was precisely controlled by the reaction duration based on Faraday's law with an assumption that the full amount of electricity due to the current was used for the electrochemical de-intercalation of Li. Typically, a 100-mg sample was charged with $I$ = 0.1 mA for 137, 178, 241, and 274 hours to obtain the $x$ = 0.50, 0.35, 0.12, and 0.0 (*i.e.* $CoO_2$) phases, respectively.

Since high-valent cobalt oxides tend to experience chemical instability when exposed to atmospheric moisture, sample handling and characterization were carefully made. After the electrochemical procedure, the samples were washed with anhydrous dimethyl carbonate in an argon-filled glovebox and then encapsulated to prevent exposure to air. XRPD analysis was carried out for electrochemically treated samples, which were set in an airtight sample holder filled with argon gas. The Li content ($x$) in the pristine and electrochemically treated samples was determined by means of inductively coupled plasma-atomic emission spectroscopy (ICP-AES). Magnetic susceptibility ($\chi$) was measured using a SQUID magnetometer (MPMS-XL; Quantum Design) in a temperature range between 2 and 300 K under a magnetic field of $H = 10$ kOe. An as-encapsulated sample was put in a cryostat for magnetic measurements: the contribution from the capsule to the magnetic data was accordingly subtracted. For each Li composition, the magnetic measurements were performed on several samples to check reproducibility. Thermal behaviors were studied on some selected samples by means of differential scanning calorimetry (DSC). The measurements were carried out with commercial equipment (Diamond DSC; Perkin Elmer) in a temperature range between 104 and 283 K.

III. RESULTS

III.1. Phase stability of Li-deficient $Li_xCoO_2$ phases

The phase stability of Li-deficient $Li_xCoO_2$ has been studied extensively for its importance as a cathode material in Li-ion secondary batteries. In a commercial battery, the $LiCoO_2$ cathode is usually cycled with an upper cutoff voltage of about 4.2 V with respect to Li metal, corresponding to extraction/insertion of 0.5 Li per $LiCoO_2$. The electrochemical behavior is well established for $1.0 \geq x \geq 0.5$ [17,18], whereas it has not been fully understood for lower Li contents below 0.5. To clarify the phase evolution upon de-intercalation of Li from $LiCoO_2$, we measured the quasi-open-circuit voltage (OCV) of the $Li_xCoO_2$/Al cell as a function of $x$ (Fig. 1). In this experiment, the cell voltage was measured with a repeated sequence of having a current of 0.1 mA turned on (for 1 h) and off (for 1 h). Here, the quasi-OCV is the relaxed voltage recorded when the current is off. It is considered that OCV is closely related to the chemical potential of the $Li_xCoO_2$ cathode. With decreasing $x$, the OCV value increases in a nonlinear manner, implying that the structural phase diagram of $Li_xCoO_2$ is somewhat complex.

To interpret the OCV data, the $x$-derivative of OCV, *i.e. dV/dx*, is plotted against $x$ (Fig. 2). A steep decrease in *dV/dx* indicates a voltage plateau caused by two-phase coexistence. In Fig. 2, we see five prominent dips in the *dV/dx–x* plot at $x = 0.94 – 0.75$, $\approx 0.55$, $\approx 0.48$, $\approx 0.36$, and $0.25 – 0.12$. The first dip at $x = 0.94 – 0.75$ corresponds to the biphasic regime that has been widely recognized [17-19]. The second and third dips ($x \approx 0.55$ and $\approx 0.48$, respectively) are located in the vicinity of $x = 1/2$, at which the monoclinic phase with Li/vacancy ordered structure appears [20]. It is thus likely that these anomalies are triggered by a strong tendency for Li/vacancy (1:1) ordering. The fourth dip at $x \approx 0.36$ is seen just above a fractional composition of $x = 1/3$. One may anticipate that it is also related to Li/vacancy (1:2) ordering as predicted by first-principles calculations, although a recent structural study evidenced no signature of Li/vacancy ordering in the $x = 0.35$ sample [21]. We note that, as we will see later, the magnetic property clearly changes around this composition. The fifth dip at $x = 0.25 – 0.12$ is inexplicable due to its broadened feature. With this result, it is not possible to determine the position of phase boundaries without ambiguity. In fact, the electrochemical behavior below $x \approx 1/3$ is controversial among previous literatures [18,22-24].

Based on the present OCV experiment, combined with *ex-situ* XRPD analysis on samples with several Li compositions, the structural phase diagram of the $Li_xCoO_2$ system has been constructed (Fig. 2). There exist seven distinct phases in the $Li_xCoO_2$ system. These phases are accordingly called as O3-$R_1 \sim R_4$, O3-M, H1-3, and O1, based on their crystallographic features (see below). Our phase diagram is in a good agreement with that previously reported, especially for the Li-rich compositions [17-19,22]. For $x < 0.5$, on the other hand, several new aspects can be pointed out. (1) A biphasic region could exist at $x = 0.35 – 0.40$ (*i.e.* the fourth dip in the *dV/dx–x* plot). It has not been reported in previous works. (2) The H1-3 phase forms within a narrow range of $x$, probably only at $x = 0.12$. This result does not agree with the first principles calculations by Van der Ven *et al.* [25], who claimed that the H1-3 structure may be stable for Li contents between $x = 0.12$ and 0.19. (3) The O1 structure appears only at $x = 0$. A well-defined phase does not exist for $0 < x < 0.12$.

Polycrystalline $Li_xCoO_2$ samples of $x = 0$ (*i.e.* $CoO_2$), 0.12, 0.35, 0.40, 0.50, 0.60,

0.67, 0.70, and 1.0 (*i.e.* pristine $LiCoO_2$) were synthesized. XRPD patterns for the samples are shown in Fig. 3. As anticipated from the phase diagram, all the samples are of single-phase. Sharp diffraction peaks throughout the XRPD patterns ensure that our $Li_xCoO_2$ and $CoO_2$ samples are chemically homogenous with good crystallinity. For $x$ = 0.35, 0.40, 0.60, 0.67, 0.70, and $LiCoO_2$, diffraction peaks are readily indexed based on rhombohedral space group *R*-3*m*. These samples crystallize in a so-called O3-type structure, in which Li ions occupy an octahedral site with three $CoO_2$ layers per unit cell (= O3-R phase) [26,27]. The $x$ = 0.5 phase also possesses the O3-type structure, but it belongs to a monoclinic system of space group *P*2/*m* (= O3-M phase) due to Li/vacancy (1:1) ordering [20]. On the other hand, layer-stacking sequences of $CoO_2$ and $x$ = 0.12 are totally different. The $CoO_2$ phase crystallizes in a hexagonal structure of space group *P*-3*m*1 containing a single $CoO_2$ layer only per unit cell (= O1 phase) [22,27]. The crystal of $x$ = 0.12 is reported to consist of alternate stacking of a Li-intercalated O3-type block (as in $LiCoO_2$) and a Li-free O1-type block (as in $CoO_2$), leading to a six-$CoO_2$-layer unit cell that is called "H1-3" [23,24]. For all the samples, Li content ($x$) determined by ICP-AES, lattice parameters, and interlayer distance are summarized in Table I. The actual Li contents are in excellent agreement with the nominal values, indicating that the full amount of electricity due to the current was used for Li de-intercalation from $LiCoO_2$. The lattice parameters of our samples are consistent with those in previous literatures [20-24].

III.2. Magnetic properties

Figure 4 shows the dependence of magnetic susceptibility ($\chi$) on temperature for samples with lower Li contents, $x$ = 0 ($CoO_2$), 0.12, 0.35, and 0.40. The susceptibility of $CoO_2$ (red symbols in Fig. 4) is nearly constant in a temperature range between 50 and 300 K, and it rapidly increases below 50 K. Although the $\chi - T$ curves of the $x$ = 0.12, 0.35 and 0.40 samples look similar to that of $CoO_2$, the magnetic behavior is clearly different among these samples, as exemplified by the normalized $\chi - T$ plots shown in Fig. 5. For $CoO_2$, the $\chi$ value slightly increases as temperature decreases until the upturn starts to grow, while the susceptibility of $x$ = 0.35 (purple symbols) decreases with decreasing temperature and reaches a broad minimum at around 100 K. The positive slope above 100 K suggests the existence of a broad peak at high temperatures. For $x$ = 0.40 (orange symbols), a small hump is seen at 30 K, suggesting the onset of magnetic ordering. Furthermore, this sample shows a more prominent upturn than the other three samples. Also, one may notice that the normalized $\chi(T)$ for $x = 0.12$ (blue symbols) perfectly coincides with that for $CoO_2$ at high temperatures, while it deviates slightly below ~100 K. The deviation was well reproducible, although it is very small.

In Fig. 6, $\chi - T$ curves for samples with the higher Li contents, $x = 0.50, 0.60, 0.67, 0.70$, and $1.0$ ($LiCoO_2$) are presented. The susceptibility of pristine $LiCoO_2$ is small in magnitude and little dependent on temperature, as the constituent $Co^{III}$ is in nonmagnetic low-spin state ($S = 0$). This result is in good agreement with those previously reported [28]. On the other hand, the Li-deficient samples exhibit complicated magnetic behaviors. For $x = 0.67$ and $0.70$, the $\chi$ value slightly increases with lowering temperature and then suddenly decreases at $T_t = 175 \sim 185$ K. The magnetic anomaly involves temperature hysteresis of $\Delta T = 4$ K between the heating and cooling curves. The $x = 0.50$ sample also shows a magnetic jump at about 175 K, as reported previously [29-31]. However, the feature is somewhat different from that of $x = 0.67$ and $0.70$: the hysteresis width is much larger for $x = 0.50$ ($\Delta T \approx 20$ K) than for $x = 0.67$ and $0.70$. Importantly, it appears that the magnetic anomaly at $T_t \approx 175$ K is absent in the sample with an intermediate Li composition, *i.e.* $x = 0.60$. Instead of the remarkable magnetic jump, the $x = 0.60$ sample exhibits a tiny kink at $T \approx 210$ K. These facts suggest that the magnetic anomaly in $x = 0.50$ is inherently different from that in $x = 0.67$ and $0.70$.

To clarify the nature of this magnetic anomaly, DSC curves were recorded for the $x = 0.50$ and $0.67$ samples (Fig. 7). Both in the two samples, latent heat is clearly observed at the magnetic anomaly point ($T_t = 175 \sim 185$ K), indicating that the anomaly is triggered by a first-order phase transition. For $x = 0.67$, both the endothermic and exothermic peaks appear in the heating and cooling curves, respectively, whereas the exothermic peak is hardly seen in the $x = 0.50$ sample. This is consistent with the fact that the magnetic jump is significantly broadened upon cooling in this sample (see green curves in Fig. 6). The latent heat $\Delta H$ is estimated from the endothermic peak area to be 82.2 and 272 J/mol for $x = 0.50$ and $0.67$, respectively. Thus, the value of $\Delta H$ is three times larger for $x = 0.67$ than $x = 0.50$. Mukai *et al.* reported muon-spin spectroscopy experiments on Li-deficient $Li_xCoO_2$ and suggested that the transition at $T_t \approx 175$ K is not magnetic but originated from either charge ordering or a change in the spin state [30]. Previous transport measurements by Ménétrier *et al.* revealed [19] that the $x = 0.70$ phase exhibits a rapid decrease in electrical conductivity below $T_t$, indicating a reduction of the carrier density at low temperatures. Taking into account these results, as well as the fact that the magnetic anomaly appears only in the vicinity of fractional Li contents, it is reasonable that the transitions are likely to be triggered by charge ordering.

The $\chi - T$ plots were fitted with the following formula:
$$\chi = \chi_0 + C/(T - \Theta) \quad (1)$$
where $\chi_0$, $C$, and $\Theta$ denote a constant susceptibility, the Curie constant, and the Weiss temperature, respectively. For $x$ = 0.50, 0.60, 0.67, and 0.70, least-square fits were carried out in a limited temperature range between 2 and 150 K, since the $\chi - T$ curves deviate from Eq. (1) due to the existence of magnetic anomalies. Also, the data of the $x$ = 0.35 sample was fitted only below 100 K, as the $\chi$ value gradually increases at elevated temperatures. The Weiss temperature $\Theta$ is always negative and small in magnitude (= $-1 \sim -5$ K), being independent of the Li content ($x$). In Fig. 8(a), $\chi_0$ is plotted as a function of $x$. The magnitude of $\chi_0$ linearly increases with decreasing $x$, except for $x$ = 0.50, 0.67, and 0.70, at which $\chi_0$ is reduced due to the presence of the magnetic jump at $T_t \approx 175$ K. It has been reported [19] that pristine $LiCoO_2$ is a band insulator, and hole doping through Li de-intercalation leads to metallic conductivity. The relatively large $\chi_0$ in Li-deficient $Li_xCoO_2$ is thus attributed to a Pauli-paramagnetic component. The increase in $\chi_0$ with decreasing $x$ implies the enhancement in the density of states at the Fermi level [$D(\varepsilon_F)$]. The $D(\varepsilon_F)$ value is calculated at 13 electrons /eV for $CoO_2$, assuming that the difference in magnitude of $\chi_0$ between $CoO_2$ and $LiCoO_2$ phases corresponds to the Pauli-paramagnetic contribution [14,15].

From the $C$ value, the effective magnetic moment $\mu_{eff}$ is readily calculated and plotted in Fig. 8(b) as a function of $x$. In this figure, blue circles denote $\mu_{eff}$ per Co (*i.e.* all the cobalt atoms are considered equivalent), and red squares are $\mu_{eff}$ per $Co^{IV}$, under the assumption that only $Co^{IV}$ spins contribute and all other $Co^{III}$ are nonmagnetic with $S$ = 0. It can be seen that the magnitude of $\mu_{eff}$ gradually increases and saturates with decreasing $x$, then it drops abruptly when the Li content is smaller than $x$ = 0.40. The $\mu_{eff}$ value for $x \leq 0.35$ is indeed comparable to that for pristine $LiCoO_2$ which contains nonmagnetic $Co^{III}$ only. We thus interpret that the small effective moment of the $x \leq$ 0.35 samples that gives the low-temperature upturn is due to an extrinsic cause, *e.g.* lattice defects. This is supported by our recent $^{59}Co$-NMR/NQR experiment in which the Curie term is not seen in the Knight shift for $x \leq 0.35$ [16]. Assuming that the upturn is attributed to $S$ = 1/2 localized spins, their concentration is estimated at 0.5 ~ 0.8%. The magnetism of $Li_xCoO_2$ with $x \leq 0.35$ is featured with a temperature-independent susceptibility with a relatively large value for $\chi_0$, strongly suggesting that the compounds are Pauli-paramagnetic metals with itinerant electrons. The similarity in the $\mu_{eff}$ value between the $x \leq 0.35$ samples and pristine $LiCoO_2$ implies that the concentration of lattice defects has remained almost unchanged even after the electrochemical oxidation procedure. Thus, the interpretation that the relatively large $\mu_{eff}$ for $0.70 \geq x \geq 0.40$ also stems from an extrinsic source is inappropriate. It is more reasonable to conclude that the magnetic moment intrinsically forms in this composition range: *i.e.* $Li_xCoO_2$ behaves as a "Curie-Weiss metal", although the magnitude of $\mu_{eff}$ (= $0.28 \sim 0.35$ $\mu_B/Co^{IV}$) is somewhat small.

## IV. DISCUSSION

IV-1. Electronic phase diagram of the $Li_xCoO_2$ system

The present study has evidenced that the electronic state of $Li_xCoO_2$ is sensitively influenced by the Li content, $x$. A distinct change in the magnetic behavior is found to take place at a critical Li content, $x_c = 0.35 \sim 0.40$: the magnetism looks like of Curie-Weiss type for $x \geq 0.40$, while it is paramagnetic with relatively large $\chi_0$ for $x \leq 0.35$. It is important to note that a similar critical point is seen in the $Na_xCoO_2$ system [7]. The similarity implies that the essential physics may be identical in the both systems. In $Na_xCoO_2$, the electronic phase diagram is divided into two regimes at $x_c \approx 0.5$. The transport properties in the paramagnetic regime ($x < 0.5$) are more conventional than those in the Curie-Weiss regime ($x > 0.5$) [7]. In the latter, thermoelectric power is greatly enhanced by a large spin-entropy component [32]. The properties in the Curie-Weiss regime are anomalous from the viewpoint of a conventional metal. Note that unusually large thermoelectric power is also reported for $Li_xCoO_2$ in the "Curie-Weiss type" regime, *i.e.* $S_{300K} = 75$ $\mu V / K$ for $x = 0.7$ [19].

Recently, we performed $^{59}$Co-NMR/NQR observations on $CoO_2$ ($x = 0.0$) and $Li_xCoO_2$ ($x = 0.12 – 0.35$) [16]. Our $^{59}$Co-NMR/NQR studies revealed a complex nature in the electronic phase diagram in the low Li content regime. It was found that antiferromagnetic-like fluctuations develop and a crossover to a Fermi-liquid regime occurs below a characteristic temperature $T^*$, when the Li content is smaller than $x = 0.35$. Remarkably, $T^*$ is found to decrease from ~50 K for $x = 0.25$ to ~7 K for $x = 0.12$, indicating that a sample with smaller $x$ is closer to magnetic instability. Nevertheless, $CoO_2$, the $x = 0$ end member is a conventional metal that well conforms to the Fermi liquid theory. The $^{59}$Co-NMR/NQR results demonstrate that the properties of the $CoO_2$, $x = 0.12$, and 0.35 phases are obviously dissimilar in terms of spin correlations. Thus, the slightly different magnetic behaviors exemplified in Fig. 5 would be attributed to the change in the spin dynamics. It should be noted that our NMR/NQR results are not in agreement with those of de Vaulx *et al.* (Ref. 33), who claimed that $CoO_2$ is an itinerant metal with clear signs of strong electron correlations.

On the basis of the dc-magnetic susceptibility data, together with the $^{59}$Co-NMR/NQR results, the electronic phase diagram of the $Li_xCoO_2$ system has been established (Fig. 9). The most prominent feature is a rich variety in electronic properties. In fact, a magnetic critical point is found to exist between $x = 0.35$ and 0.40, which separates a Pauli-paramagnetic and a Curie-Weiss metals. In the Pauli-paramagnetic regime ($x < x_c$), a magnetic crossover takes place at the characteristic temperature $T^*$: antiferromagnetic (AF) fluctuations develop above $T^*$, and $T^*$ tends to be lowered with decreasing $x$. Although the magnetic crossover could not be observed for $x = 0.35$ below 150 K, the susceptibility data (Fig. 5) suggest that electron correlations would develop at higher temperatures. This means that $Li_xCoO_2$ in the low Li content regime is regarded as an itinerant metal involving electron correlations, and the spin fluctuations are enhanced when approaching $x = 0$. This is consistent with the picture that members of $A_xCoO_2$ ($A$ = Li, Na) with small $x$ can be viewed as a doped spin-1/2 system. On the other hand, $CoO_2$, the $x = 0$ end member is a conventional metal in the whole temperature range studied. The disappearance of electron correlations in $CoO_2$ is somewhat surprising. The weakly correlated nature in $CoO_2$ is believed to originate from the abrupt change in the crystal structure. The crystal of $CoO_2$ is less anisotropic, since there is no "spacer" layer between two adjacent $CoO_2$ blocks. A more three dimensional electronic structure is likely to suppress the spin fluctuations in $CoO_2$.

For $x > x_c$, the effective magnetic moment is now significant: it is considered that $Li_xCoO_2$ behaves as a "Curie-Weiss metal". Like $Na_xCoO_2$, large thermoelectric power and metallic conductivity are simultaneously observed in this regime [19]: the properties are thus more anomalous than those in the Pauli-paramagnetic regime ($x < x_c$). Another characteristic feature in this regime is the appearance of various phase transitions, depending on the Li content. For $x = 0.40$, a susceptibility hump is seen at 30 K, suggesting the onset of static AF order. The possibility that the hump is due to magnetic impurities (such as $Co_3O_4$) can be ruled out, since the hump was seen in samples only with $x = 0.40$ and it never appeared in other Li compositions. Details in this magnetic behavior are still unclear, and further investigations are necessary. Except for $x = 0.40$, static magnetic order seems to be absent in the $Li_xCoO_2$ system. This is in contrast to the $Na_xCoO_2$ system where AF spin arrangement is detected at $x = 0.5$ [7] and $\geq 0.75$ [5,34].

On the other hand, first-order phase transitions are observed at $x = 0.50$, 0.67, and 0.70. The transitions are likely to involve charge ordering in the vicinity of fractional Li contents at $x = 0.50$ (= 1/2) and 0.67 (= 2/3). From the magnitude of latent heat ($\Delta H$), the transition entropy is readily estimated: $\Delta S = 0.47$ and 1.49 J/K mol for $x = 0.50$ and 0.67, respectively. These values are much smaller than the theoretical values of "mixing entropy", *i.e.* $\Delta S_{mixing} = -R\,(1/2 \ln 1/2 + 1/2 \ln 1/2) = 5.76$ J/K mol and $-R\,(1/3 \ln 1/3 + 2/3 \ln 2/3) = 5.29$ J/K mol for $x = 1/2$ and 2/3, respectively [35]. The smaller $\Delta S$ values suggest that the charge ordering may be incomplete: cobalt species separate into two states with decimal valence numbers (= charge disproportionation, *e.g.* $2Co^{+3.5} \to Co^{+3.5+\delta} + Co^{+3.5-\delta}$), or only a part of carriers are localized below $T_t$. These speculations are in good agreement with the fact that the $\chi_0$ value (*i.e.* Pauli-paramagnetic component) is finite in the $x = 0.50$, 0.67, and 0.70 samples. Indeed, incomplete or partial localization of electrons has also been reported for the $Na_xCoO_2$ system [36,37]. The magnetic anomaly is also detected at $x = 0.60$. We interpret that the anomaly of the $x = 0.60$ sample is not associated with the transition observed for $x = 0.50$, 0.67, and 0.70, because the behavior is apparently different in terms of magnitude and temperature. The origin of the anomaly for $x = 0.60$ is unclear and open to dispute.

With Li contents close to $x = 1$, $Li_xCoO_2$ has been regarded as a band insulator [19]. A recent work by Ménétrier *et al.* [38] has demonstrated that highly stoichiometric $Li_1CoO_2$ exhibits a very early insulator to metal transition upon Li de-intercalation not at $x = 0.94$ but at $x = 1-\varepsilon$ ($\varepsilon \ll 1$). This result implies that the lowest boundary of the O3-$R_1$ phase is highly sensitive to the concentration of crystal defects.

There are earlier reports on the magnetism and electronic structure of $CoO_2$ and $Li_xCoO_2$. Mukai *et al.* investigated the magnetic phase diagram of $Li_xCoO_2$ ($x = 0.1 - 1.0$) by means of muon-spin spectroscopy and susceptibility measurements [30]. We emphasize that several important aspects are missing in their phase diagram: they reported neither the magnetic critical point at $x_c = 0.35 \sim 0.40$ nor the development of spin fluctuations in the low $x$ regime. Also, the authors did not recognize the abrupt change in the electronic structure between $Li_xCoO_2$ and $CoO_2$. Due to the lack of these aspects, the phase diagram given in Ref. [30] is much simpler than ours. On the other hand, Hertz et al. reported magnetic properties of $Li_xCoO_2$ with Li contents $0.5 < x < 1.0$ [39]. They found that all of Li-deficient samples show a Curie-Weiss behavior, indicating the existence of local cobalt moments in their samples. This finding is in good agreement with our observations in the present study, although the $\mu_{eff}$ values in Ref. 39 are rather larger than ours. They also claimed that in samples with $x \approx 0.7$ the magnitude of $\mu_{eff}$ per $Co^{IV}$ is consistent with the theoretical spin-only value of low-spin $Co^{IV}$, but this conclusion is based on a miscalculation of $\mu_{eff}$, as pointed out in Ref. [38].

IV-2. Comparison to the $Na_xCoO_2$ system

Despite the similarity in $Li_xCoO_2$ and $Na_xCoO_2$ with respect to the magnetic critical point, two quantitative differences can be pointed out between the two systems. First, the critical point $x_c$ is smaller in $Li_xCoO_2$ than $Na_xCoO_2$, i.e. $x_c = 0.35 \sim 0.40$ and $\approx 0.5$ in the former and the latter, respectively. A recent investigation by Yoshizumi et al. demonstrated [40] that the critical point $x_c$ in $Na_xCoO_2$ lies between 0.58 and 0.59. Yokoi et al. also claimed [41] that the critical point is situated at about 0.60. In $Li_xCoO_2$, on the other hand, a low-temperature upturn is still prominent around this composition (see the data of $x = 0.50$ and 0.60 in Fig. 6). Second, $\mu_{eff}$ is different in magnitude between $Li_xCoO_2$ and $Na_xCoO_2$. In the former, the magnitude of $\mu_{eff}$ (= 0.28 ~ 0.35 $\mu_B/Co^{IV}$) is much smaller than the theoretical spin-only value of $Co^{IV}$ (= 1.73 $\mu_B$), while in the latter the $\mu_{eff}$ value is consistent with a spin-1/2 local-moment population equal to the $Co^{IV}$ concentration [7]. Since these two systems contain the triangular $CoO_2$ block in common, these differences are believed to originate from modifications in the crystal structure.

Yoshizumi et al. proposed a possible interpretation for the critical point ($x_c$) in $Na_xCoO_2$ [40]. According to their argument, the critical point corresponds to a characteristic Na content at which the topology of the Fermi surfaces (FS) substantially changes, leading to large modifications in the electronic properties. Previous band calculations indicated the existence of a dip structure around the Γ point in the $a_{1g}$ band [42,43]. Then, it is likely that at $x = x_c$ the Fermi level touches the bottom of the dip exactly at the Γ point, as depicted in Fig. 5 in Ref. [40]. One expects that for $x < x_c$ only a single cylindrical FS exists, while for $x > x_c$ an additional small electron pocket should appear around the Γ point. Thus, the anomalous electronic properties in the Curie-Weiss regime are attributed to the emergence of this small electron pocket. Based on this scenario, it is reasonable to consider that the $x_c$ value depends on the crystal structure, since the band structure is sensitively related to the local environment of $CoO_6$ octahedra. We thus speculate that the different $x_c$ values in $Li_xCoO_2$ and $Na_xCoO_2$ are due to slight changes in the local structure in the $CoO_2$ block. The difference in the local structure was indeed reported: it was found that $CoO_6$ octahedra in $Li_xCoO_2$ are less distorted than those in $Na_xCoO_2$ [44]. Theoretical studies on the electronic structure of $Li_xCoO_2$ are highly desirable, in order to prove the above argument.

For $Li_xCoO_2$, the $\mu_{eff}$ value in the Curie-Weiss regime is much smaller than the theoretical spin-only value of $Co^{IV}$. Also, the Weiss temperature is small in magnitude, *i.e.* Θ = −1 ~ −5 K, being in sharp contrast to large negative values in $Na_xCoO_2$: Θ = −156 K and −99 K for $x$ = 0.59 and 0.70, respectively [40]. These facts imply that the electron correlation effect is weaker in $Li_xCoO_2$. Nevertheless, it should be noted that $Li_xCoO_2$ also exhibits large thermoelectric power and metallic conductivity simultaneously [19]: the property is obviously unusual, although the electron correlation effect is less prominent in $Li_xCoO_2$. We believe that physics in the "Curie-Weiss metal" in $Li_xCoO_2$ is essentially identical to that in $Na_xCoO_2$. Then, a question arises: what is the origin of the quantitative differences in the magnetic properties between the two systems? We suggest that the dimensionality of the electronic structure plays an important role. The crystal of $Li_xCoO_2$ is more three-dimensional than that of $Na_xCoO_2$ due to its shorter interlayer Co-Co distance, *i.e.* $d_{Co-Co}$ = 4.7 – 4.8 Å and 5.4 – 5.5 Å for the former and the latter, respectively. It is likely that the weakened electron correlation in $Li_xCoO_2$ is a consequence of the more three dimensional electronic structure. The importance of the dimensionality is also suggested by the experimental fact that the $CoO_2$ phase does not show any indication of electron correlations [16]. Theoretical investigations are thus urgent to elucidate how the $\mu_{eff}$ and Θ values vary along with the electron correlation effects.

Finally, we comment on the absence of any electronic phase transition in the vicinity of $x$ = 2/3 in $Na_xCoO_2$, contrary to general expectation for a triangular lattice. Chou *et al.* reported [11] that there is a strong tendency of Na ion ordering at $x$ = 0.71 with a large superstructure consisting of 12 unit cells. Thus, it is suggested that such a stable Na-ion superstructure still survives around $x$ = 2/3, and a Coulomb potential due to long-range Na-ion order highly prevents the formation of charge ordering in the $CoO_2$ block. This is not the case of Li$_x$CoO$_2$, in which no indication of Li ion ordering is evidenced around $x = 2/3$ [21]. From these facts, we think that the CoO$_2$ block in Na$_x$CoO$_2$ is more strongly perturbed by the neighboring Na ion block. In other words, Li$_x$CoO$_2$ may be a more appropriate system than Na$_x$CoO$_2$ for investigations on true physics in the triangular CoO$_2$ lattice.

## V. CONCLUSIONS

The magnetic properties of the layered cobalt oxide system, Li$_x$CoO$_2$, were systematically investigated in the whole range of Li composition, $0 \leq x \leq 1$. Based on dc-magnetic susceptibility data, combined with results of $^{59}$Co-NMR/NQR observations [16], the electronic phase diagram of Li$_x$CoO$_2$ was established. It was found that the phase diagram of Li$_x$CoO$_2$ is complex, and the electronic properties are sensitively influenced by the Li content ($x$). As in the related material Na$_x$CoO$_2$, a magnetic critical point was found to exist between $x = 0.35$ and $0.40$, which separates a Pauli-paramagnetic and a Curie-Weiss metals. The similarity in the magnetic behaviors implies that the essential physics may be identical in both the Li$_x$CoO$_2$ and Na$_x$CoO$_2$ systems. In the Pauli-paramagnetic regime ($x \leq 0.35$), the antiferromagnetic (AF) spin correlations systematically increase with decreasing $x$. Nevertheless, CoO$_2$, the $x = 0$ end member is a non-correlated metal in the whole temperature range studied. The disappearance of the electron correlations in CoO$_2$ is believed to originate from the abrupt change in the crystal structure. In the Curie-Weiss regime ($x \geq 0.40$), on the other hand, various phase transitions were observed. For $x = 0.40$, a susceptibility hump is seen at 30 K, suggesting the onset of static AF order. A magnetic jump, which is likely triggered by charge ordering, was clearly observed at $T_t \approx 175$ K in samples with $x = 0.50$ (= 1/2) and 0.67 (= 2/3), while only a tiny kink appears at $T \approx 210$ K in the sample with an intermediate Li composition, $x = 0.60$. Despite the similarity in Li$_x$CoO$_2$ and Na$_x$CoO$_2$ with respect to the magnetic critical point, quantitative differences were found between the two systems. It is suggested that the differences are caused by modifications in the crystal structure.

## ACKNOWLEDGMENTS

The authors thank G.-q. Zheng, S. Kawasaki, T. Tohyama, and W. Koshibae for their fruitful discussion and comments. Also, S. Nakamura (of the Center for Advanced Materials Analysis Technical Department, Tokyo Institute of Technology) is acknowledged for ICP-AES analysis. The present work was supported by Grants-in-aid for Scientific Research (Contracts 16740194 and 19740201) from the Japan Society for the Promotion of Science. H.Y. acknowledges financial support from Tekes (No. 1726/31/07) and M. K. from the Academy of Finland (No. 110433).

REFEFENCES

[1] C. Fouassier, G. Matejka, J.-M. Reau, and P. Hagenmuller, J. Solid State Chem. **6**, 532 (1973).

[2] I. Terasaki, Y. Sasago, and K. Uchinokura, Phys. Rev. B **56**, R12685 (1997).

[3] T. Motohashi, E. Naujalis, R. Ueda, K. Isawa, M. Karppinen, and H. Yamauchi, Appl. Phys. Lett. **79**, 1480 (2001); T. Motohashi, M. Karppinen, and H. Yamauchi, in Oxide Thermoelectrics, Research Signpost, India, 2002, pp. 73-81.

[4] M. Lee, L. Viciu, L. Li, Y. Wang, M.L. Foo, S. Watauchi, R.A. Pascal Jr., R.J. Cava, and N.P. Ong, Nature Mater. **5**, 537 (2006).

[5] T. Motohashi, R. Ueda, E. Naujalis, T. Tojo, I. Terasaki, T. Atake, M. Karppinen, and H. Yamauchi, Phys. Rev. B **67**, 064406 (2003).

[6] K. Takada, H. Sakurai, E. Takayama-Muromachi, F. Izumi, R.A. Dilanian, T. Sasaki, Nature (London) **422**, 53 (2003).

[7] M. L. Foo, Y. Wang, S. Watauchi, H.W. Zandbergen, T. He, R.J. Cava, and N.P. Ong, Phys. Rev. Lett. **92**, 247001 (2004).

[8] M. Karppinen, I. Asako, T. Motohashi, and H. Yamauchi, Phys. Rev. B **71**, 092105 (2005).

[9] H. W. Zandbergen, M. L. Foo, Q. Xu, V. Kumar, and R.J. Cava, Phys. Rev. B **70**, 024101 (2004).

[10] G. J. Shu, A. Prodi, S.Y. Chu, Y.S. Lee, H.S. Sheu, and F.C. Chou, Phys. Rev. B **76**, 184115 (2007).

[11] F.C. Chou, M.-W. Chu, G.J. Shu, F.-T. Huang, W.W. Pai, H.S. Sheu, and P.A. Lee, Phys. Rev. Lett. **101**, 127404 (2008).

[12] K. Mizushima, P.C. Jones, P.J. Wiseman, and J.B. Goodenough, Mater. Res. Bull. **15**, 783 (1980).

[13] S. Miyazaki, S. Kikkawa, and M. Koizumi, Synthetic Metals **6**, 211 (1983).

[14] T. Motohashi, Y. Katsumata, T. Ono, R. Kanno, M. Karppinen, and H. Yamauchi, Chem. Mater. **19**, 5063 (2007).

[15] T. Motohashi, T. Ono, Y. Katsumata, R. Kanno, M. Karppinen, and H. Yamauchi, J. Appl. Phys. **103**, 07C902 (2008).

[16] S. Kawasaki, T. Motohashi, K. Shimada, T. Ono, R. Kanno, M. Karppinen, H. Yamauchi, and G.-q. Zheng, Physical Review B **79**, 220514(R) (2009).

[17] J.N. Reimers and J.R. Dahn, J. Electrochem. Soc. **139**, 2091 (1992).

[18] T. Ohzuku and A. Ueda, J. Electrochem. Soc. **141**, 2972 (1994).

[19] M. Ménétrier, I. Saadoune, S. Levasseur, and C. Delmas, J. Mater. Chem. **9**, 1135 (1999).

[20] Y. S. Horn, S. Levasseur, F. Weill, and C. Delmas, J. Electrochem. Soc. **150**, A366 (2003).

[21] Y. Takahashi, N. Kijima, K. Dokko, M. Nishizawa, I. Uchida, J. Akimoto, J. Solid State Chem. **180**, 313 (2007).

[22] G.G. Amatucci, J.M. Tarascon, and L.C. Klein, J. Electrochem. Soc. **143**, 1114 (1996).

[23] X.Q. Yang, X. Sun, and J. McBreen, Electrochem. Commun. **2**, 100 (2000).

[24] Z. Chen, Z. Lu, and J.R. Dahn, J. Electrochem. Soc. **149**, A1604 (2002).

[25] A. Van der Ven, M.K. Aydinol, G. Ceder, G. Kresse, and J. Hafner, Phys. Rev. B **58**, 2975 (1998); A. Van der Ven, M.K. Aydinol, G. Ceder, J. Electrochem. Soc. **145**, 2149 (1998).

[26] C. Delmas, C. Fouassier, and P. Hagenmuller, Physica B **99**, 81 (1980).

[27] S. Venkatraman and A. Manthiram, Chem. Mater. 14, 3907 (2002).

[28] S. Levasseur, M. Ménétrier, Y. Shao-Horn, L. Gautier, A. Audemer, G. Demazeau, A. Largeteau, and C. Delmas, Chem. Mater. **15**, 348 (2003).

[29] S. Kikkawa, S. Miyazaki, and M. Koizumi, J. Solid State Chem. **62**, 35 (1986).

[30] K. Mukai, Y. Ikedo, H. Nozaki, J. Sugiyama, K. Nishiyama, D. Andreica, A. Amato, P.L. Russo, E.J. Ansaldo, J.H. Brewer, K.H. Chow, K. Ariyoshi, and T. Ohzuku, Phys. Rev. Lett. **99**, 087601 (2007); K. Mukai, J. Sugiyama, Y. Ikedo, D. Andreica, A. Amato, J.H. Brewer, E.J. Ansaldo, P.L. Russo, K.H. Chow, K. Ariyoshi, and T. Ohzuku, J. Phys. Chem. Solid **69**, 1479 (2008).

[31] K. Miyoshi, H. Kondo, M. Miura, C. Iwai, K. Fujiwara, and J. Takeuchi, J. Phys. Conf. Series **150**, 042129 (2009).

[32] Y. Wang, N.S. Rogado, R.J. Cava, and N.P. Ong, Nature (London) **423**, 425 (2003).

[33] C. de Vaulx, M.-H. Julien, C. Berthier, S. Hébert, V. Pralong, and A. Maignan,

Phys. Rev. Lett. **98**, 246402 (2007).

[34] S.P. Bayrakci, C. Bernhard, D.P. Chen, B. Keimer, R.K. Kremer, P. Lemmens, C.T. Lin, C. Niedermayer, and J Strempfer, Phys. Rev. B **69**, 100410(R) (2004).

[35] see *e.g.*, *Atkins' Physical Chemistry 8th Revised Edition*, P. Atkins and J. de Paula (Oxford University Press, Oxford, 2006) p. 136.

[36] H. Alloul, I.R. Mukhamedshin, G. Collin, and N. Blanchard, Europhys. Lett. **82**, 17002 (2008).

[37] M.-H. Julien, C. de Vaulx, H. Mayaffre, C. Berthier, M. Horvatić, V. Simonet, J. Wooldridge, G. Balakrishnan, M.R. Lees, D.P. Chen, C.T. Lin, and P. Lejay, Phys. Rev. Lett. **100**, 096405 (2008).

[38] M. Ménétrier, D. Carlier, M. Blangero, and C. Delmas, Electrochem. Solid-State Lett. **11**, A179 (2008).

[39] J.T. Hertz, Q. Huang, T. McQueen, T. Klimczuk, J.W.G Bos, L. Viciu, and R.J. Cava, Phys. Rev. B **77**, 075119 (2008).

[40] D. Yoshizumi, Y. Muraoka, Y. Okamoto, Y. Kiuchi, J.-I. Yamaura, M. Mochizuki, M. Ogata, and Z. Hiroi, J. Phys. Soc. Jpn. **76**, 063705 (2007).

[41] M. Yokoi, T. Moyoshi, Y. Kobayashi, M. Soda, Y. Yasui, M. Sato, and K. Kakurai, J. Phys. Soc. Jpn. **74**, 3046 (2005).

[42] D.J. Singh, Phys. Rev. B **61**, 13397 (2000); D.J. Singh, Phys. Rev. B **68**, 020503(R) (2003).

[43] P. Zhang, W. Luo, M.L. Cohen, and S.G. Louie, Phys. Rev. Lett. **93**, 236402 (2004).

[44] The degree of distortion of $CoO_6$ octahedra is measured by the O-Co-O angle in the $CoO_2$ layer. The smaller the O-Co-O angle is, the more the $CoO_6$ octahedra are strongly distorted. The O-Co-O angle is larger for $Li_xCoO_2$ than for $Na_xCoO_2$, *i.e.* 85.15 deg. ($x = 0.68$ [21]) and 84.13 deg. ($x = 0.61$ [45] and 0.74 [46]) for the former and the latter, respectively.

[45] J.D. Jorgensen, M. Avdeev, D.G. Hinks, J.C. Burley, and S. Short, Phys. Rev. B **68**, 214517 (2003).

[46] L. Viciu, Q. Huang, and R.J. Cava, Phys. Rev. B **73**, 212107 (2006).

Table I. The nominal Li content, actual Li content, and crystallographic parameters of the $Li_xCoO_2$ samples.

| Nominal Li content ($x$) | Actual Li content ($x$) | Structural type | Space group | Lattice parameters | Interlayer distance $d$ |
|---|---|---|---|---|---|
| 0.0 | < 0.01 | O1 | $P$-3$m$1 | $a$ = 2.820(0) Å<br>$c$ = 4.238(1) Å | 4.24 Å |
| 0.12 | 0.12(1) | H1-3 | $R$-3$m$ | $a$ = 2.821(0) Å<br>$c$ = 27.13(0) Å | 4.52 Å |
| 0.35 | 0.35(1) | O3-$R_4$ | $R$-3$m$ | $a$ = 2.809(0) Å<br>$c$ = 14.44(0) Å | 4.81 Å |
| 0.40 | 0.39(1) | O3-$R_3$ | $R$-3$m$ | $a$ = 2.810(0) Å<br>$c$ = 14.44(0) Å | 4.81 Å |
| 0.50 | 0.49(1) | O3-M | $P$2/$m$ | $a$ = 4.866(0) Å<br>$b$ = 2.810(0) Å<br>$c$ = 5.058(0) Å<br>$\beta$ = 107.83(0)° | 4.81 Å |
| 0.60 | 0.59(1) | O3-$R_2$ | $R$-3$m$ | $a$ = 2.811(0) Å<br>$c$ = 14.35(0) Å | 4.78 Å |
| 0.67 | 0.67(1) | O3-$R_2$ | $R$-3$m$ | $a$ = 2.811(0) Å<br>$c$ = 14.31(0) Å | 4.77 Å |
| 0.70 | 0.70(1) | O3-$R_2$ | $R$-3$m$ | $a$ = 2.812(0) Å<br>$c$ = 14.28(0) Å | 4.76 Å |
| 1.0 | 0.99(1) | O3-$R_1$ | $R$-3$m$ | $a$ = 2.814(0) Å<br>$c$ = 14.05(0) Å | 4.68 Å |

Figure captions

Fig. 1 (color online).
Voltage curves of the $Li_xCoO_2$/Al electrochemical cell as lithium is de-intercalated from $LiCoO_2$. The red curve represents cell voltage under an applied current of 0.1 mA, while the blue curve is quasi-open-circuit voltage (OCV) recorded when the current is off. The Li content ($x$) of $Li_xCoO_2$ is calculated based on Faraday's law.

Fig. 2 (color online).
The structural phase diagram of the $Li_xCoO_2$ system based on the OCV experiment. In this figure, the $x$-derivative of OCV, *i.e. dV/dx*, is plotted against $x$ in order to highlight biphasic regions in the phase diagram (see main text).

Fig. 3 (color online).
X-ray powder diffraction patterns for the $Li_xCoO_2$ samples. For clarity of the figure, only five (*i.e.* $x$ = 0.0, 0.12, 0.35, 0.50, 1.0) out of the nine samples are selected.

Fig. 4 (color online).
Temperature dependence of magnetic susceptibility ($\chi$) for the $x$ = 0.0 (*i.e.* $CoO_2$), 0.12, 0.35, and 0.40 samples. For clarity of the figure, each $\chi(T)$ curve is shifted by $10^{-3}$ emu / mol Oe.

Fig. 5 (color online).
$\chi - T$ plots for the $x$ = 0.0 (*i.e.* $CoO_2$), 0.12, 0.35, and 0.40 samples. In each plot, the $\chi(T)$ values are normalized by the 300-K value.

Fig. 6 (color online).
Temperature dependence of magnetic susceptibility ($\chi$) for the $x$ = 0.50, 0.60, 0.67, 0.70, and 1.0 (*i.e.* pristine $LiCoO_2$) samples. For clarity of the figure, each $\chi(T)$ curve is shifted by $10^{-3}$ emu / mol Oe.

Fig. 7 (color online).
Differential scanning calorimetry (DSC) curves for the $x$ = 0.50 and 0.67 samples. In this experiment, the samples were first heated from 104 to 283 K, then cooled down to 104 K with a scan rate of 20 K / min.

Fig. 8 (color online).

(a) The constant susceptibility $\chi_0$ for the Li$_x$CoO$_2$ samples. (b) The effective magnetic moment $\mu_{\text{eff}}$ for the Li$_x$CoO$_2$ samples. In this figure, blue circles denote $\mu_{\text{eff}}$ per Co (*i.e.* all the cobalt atoms are considered equivalent), and red squares are $\mu_{\text{eff}}$ per Co$^{\text{IV}}$, assuming that only Co$^{\text{IV}}$ spins contribute and all other Co$^{\text{III}}$ are nonmagnetic with $S = 0$.

Fig. 9 (color online).

The electronic phase diagram of the Li$_x$CoO$_2$ system. The diagram has been constructed on the basis of the dc-magnetic susceptibility data and the $^{59}$Co-NMR/NQR results [16].

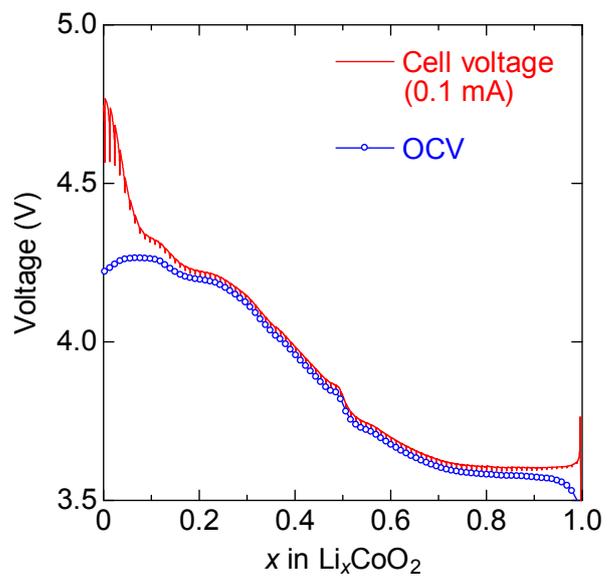

Fig. 1. T. Motohashi *et al.*

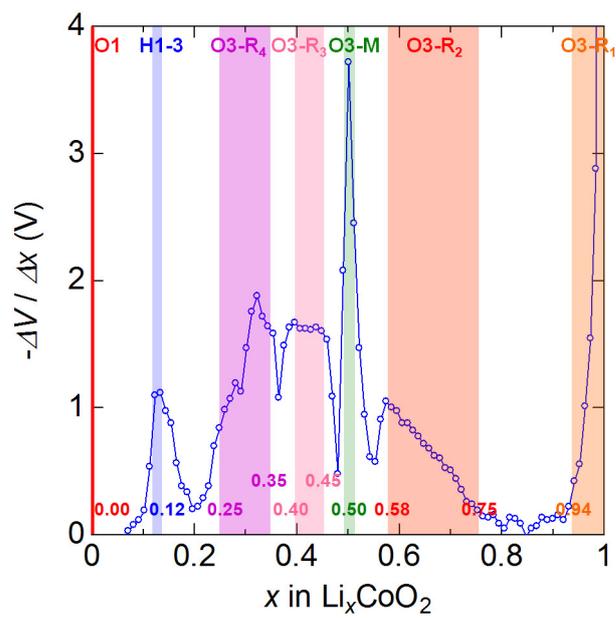

Fig. 2. T. Motohashi *et al.*

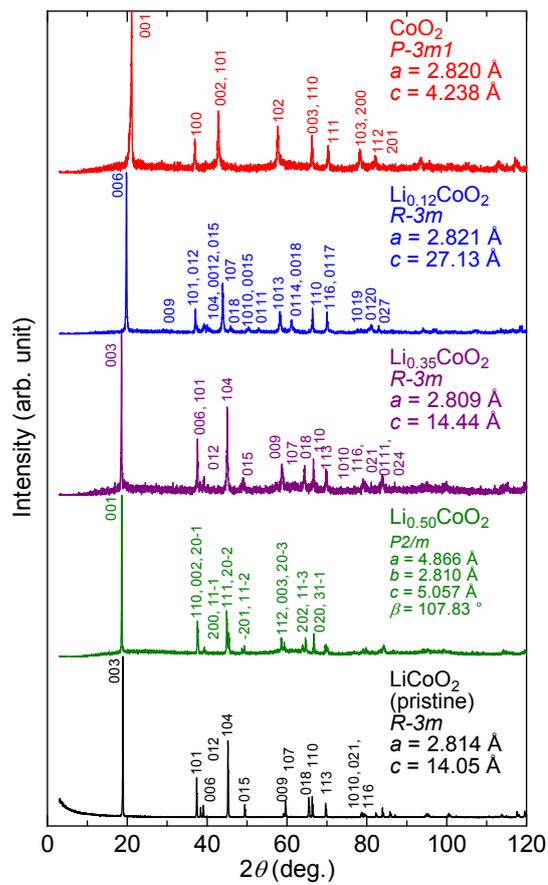

Fig. 3. T. Motohashi *et al.*

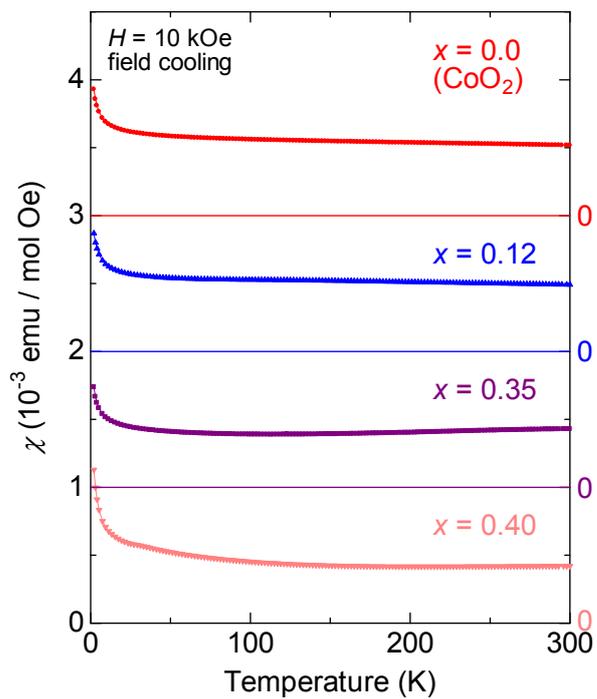

Fig. 4. T. Motohashi *et al.*

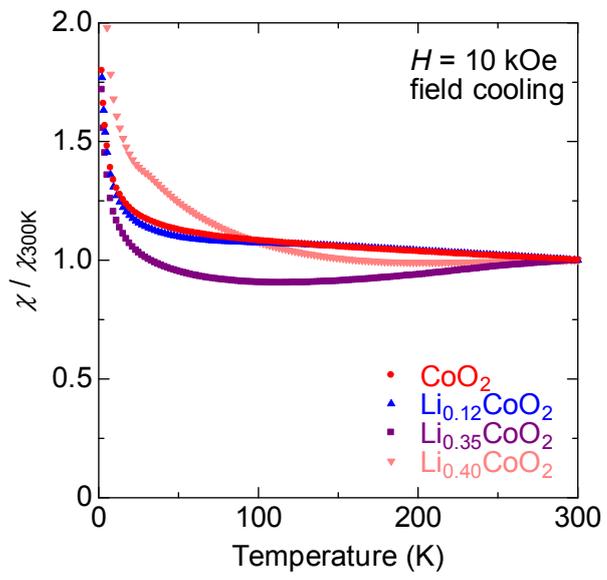

Fig. 5. T. Motohashi *et al.*

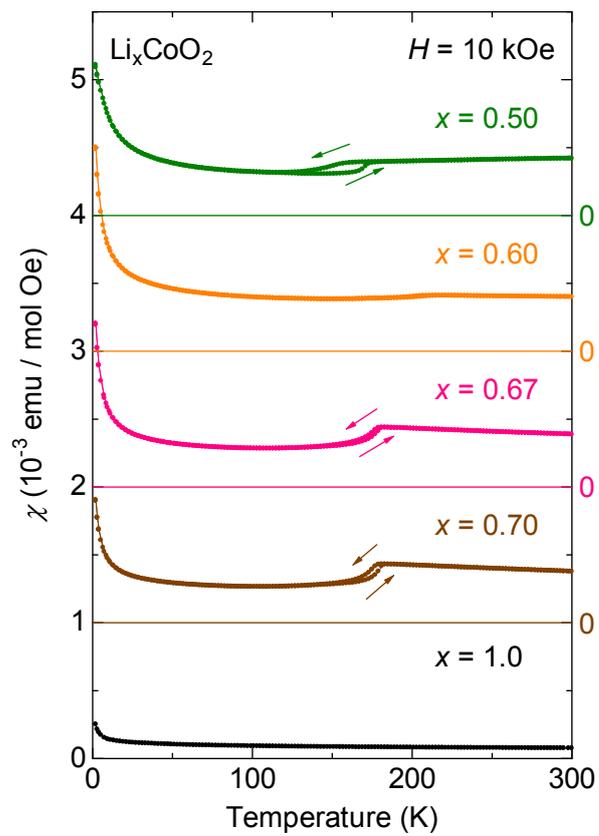

Fig. 6. T. Motohashi *et al.*

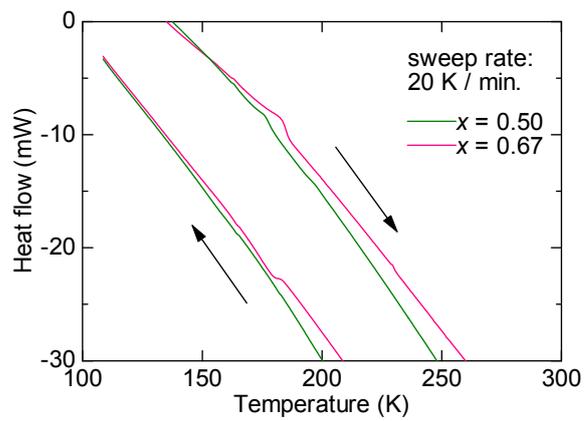

Fig. 7. T. Motohashi *et al.*

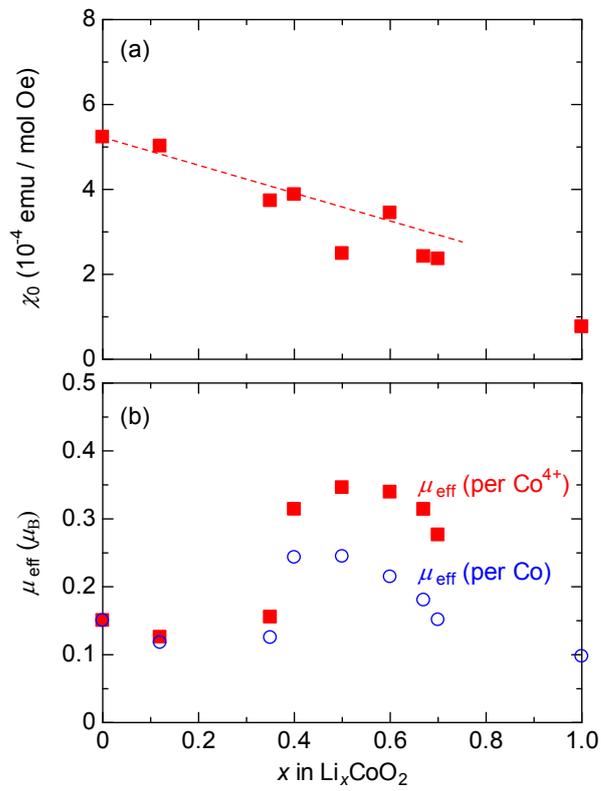

Fig. 8. T. Motohashi *et al.*

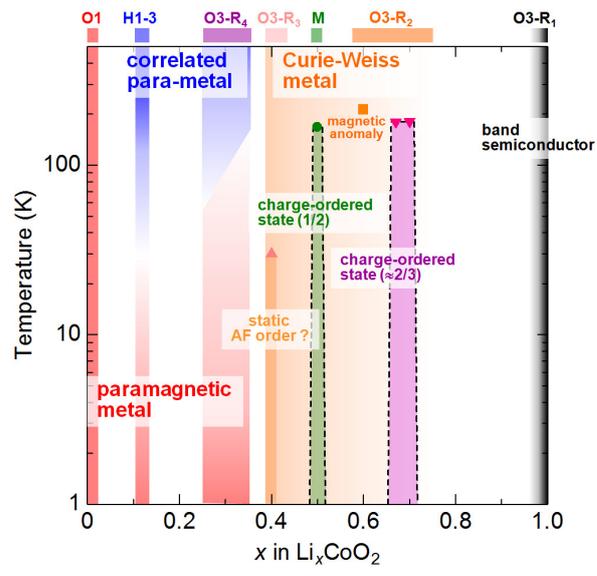

Fig. 9. T. Motohashi *et al.*